\documentstyle[12pt]{article}
\newcommand\mpl{M_{\rm P}}

\catcode`\@=11
\def\marginnote#1{}
\def\ifmath#1{\relax\ifmmode #1\else $#1$\fi}

\def\bold#1{\setbox0=\hbox{$#1$}%
     \kern-.025em\copy0\kern-\wd0
     \kern.05em\copy0\kern-\wd0
     \kern-.025em\raise.0433em\box0 }

\def\GENITEM#1;#2{\par\vskip6pt \hangafter=0 \hangindent=#1
   \Textindent{$ #2$ }\ignorespaces}

\newcount\hour
\newcount\minute
\newtoks\amorpm
\hour=\time\divide\hour by60
\minute=\time{\multiply\hour by60 \global\advance\minute by-
\hour}
\edef\standardtime{{\ifnum\hour<12 \global\amorpm={am}%
    \else\global\amorpm={pm}\advance\hour by-12 \fi
    \ifnum\hour=0 \hour=12 \fi
    \number\hour:\ifnum\minute<100\fi\number\minute\the\amorpm}}
\edef\militarytime{\number\hour:\ifnum\minute<100\fi\number\minute}
\def\draftlabel#1{{\@bsphack\if@filesw {\let\thepage\relax
  \xdef\@gtempa{\write\@auxout{\string
    \newlabel{#1}{{\@currentlabel}{\thepage}}}}}\@gtempa
    \if@nobreak \ifvmode\nobreak\fi\fi\fi\@esphack}
     \gdef\@eqnlabel{#1}}
\def\@eqnlabel{}
\def\@vacuum{}
\def\draftmarginnote#1{\marginpar{\raggedright\scriptsize\tt#1}}
\def\draft{\oddsidemargin -.5truein
        \def\@oddfoot{\sl preliminary draft \hfil
        \rm\thepage\hfil\sl\today\quad\militarytime}
        \let\@evenfoot\@oddfoot \overfullrule 3pt
        \let\label=\draftlabel
        \let\marginnote=\draftmarginnote

\def\@eqnnum{(\theequation)\rlap{\kern\marginparsep\tt\@eqnlabel}%
\global\let\@eqnlabel\@vacuum}  }
\def\preprint{\twocolumn\sloppy\flushbottom\parindent 1em
        \leftmargini 2em\leftmarginv .5em\leftmarginvi .5em
        \oddsidemargin -.5in    \evensidemargin -.5in
        \columnsep 15mm \footheight 0pt
        \textwidth 250mmin      \topmargin  -.4in
        \headheight 12pt \topskip .4in
        \textheight 175mm
        \footskip 0pt

\def\@oddhead{\thepage\hfil\addtocounter{page}{1}\thepage}
        \let\@evenhead\@oddhead \def\@oddfoot{} \def\@evenfoot{}
}
\def\titlepage{\@restonecolfalse\if@twocolumn\@restonecoltrue\o
necolumn
     \else \newpage \fi \thispagestyle{empty}\c@page\z@
        \def\thefootnote{\fnsymbol{footnote}} }
\def\endtitlepage{\if@restonecol\twocolumn \else  \fi
        \def\thefootnote{\arabic{footnote}}
        \setcounter{footnote}{0}}  
\catcode`@=12
\relax
\def\be{\begin{equation}}
\def\ee{\end{equation}}
\def\bea{\begin{eqnarray}}
\def\eea{\end{eqnarray}}
\def\simlt{\stackrel{<}{{}_\sim}}
\def\simgt{\stackrel{>}{{}_\sim}}

\def\mst11{m_{\;\widetilde{t}_{1}}}

\def\mst22{m_{\;\widetilde{t}_{2}}}
\def\mst12{m_{\;\widetilde{t}_{1,2}}}

\def\msb11{m_{\;\widetilde{b}_{1}}}
\def\msb22{m_{\;\widetilde{b}_{2}}}
\def\msb12{m_{\;\widetilde{b}_{1,2}}}

\def\mwidetilde2{\widetilde{m}^{2}}

\relax
\def\beq{\begin{equation}}
\def\eeq{\end{equation}}
\def\bea{\begin{eqnarray}}
\def\eea{\end{eqnarray}}
\def\dslash{\partial\llap{/}}
\def\hgam{{\hat \gamma}}
\def\adot{\dot{a}}
\def\addot{\ddot{a}}
\def\mdot{\dot{m}}
\def\identity{1 \hspace{-.085cm}{\rm l}}
\def\unit{\hbox to 3.3pt{\hskip1.3pt \vrule height 7pt width .4pt \hskip.7pt
\vrule height 7.85pt width .4pt \kern-2.4pt
\hrulefill \kern-3pt
\raise 4pt\hbox{\char'40}}}
\newcommand{\eq}[1]{eq.~(\ref{#1})}

\begin{document}
\input epsf

\topmargin-2.5cm
%
\begin{titlepage}
\begin{flushright}
CERN-TH/99-227\\
hep--ph/9907510 \\
\end{flushright}
\vskip 0.3in
\begin{center}
{\Large\bf Non-Thermal Production of Dangerous Relics}
\vskip 0.2cm 
{\Large\bf in the Early Universe}

\vskip .5in
{\large\bf G.F. Giudice\footnote{\baselineskip=16pt 
E-mail: {\tt Gian.Giudice@cern.ch }}}, {\large\bf I. Tkachev\footnote{
\baselineskip=16pt 
E-mail: {\tt Igor.Tkachev@cern.ch }}}
 {\bf and} 
{\large \bf A. Riotto\footnote{\baselineskip=16pt Email:
{\tt riotto@nxth04.cern.ch}}}

\vskip1cm
CERN Theory Division,

\vskip 0.2cm

CH-1211 Geneva 23, Switzerland.

\end{center}
\vskip1.3cm
\begin{center}
{\bf Abstract}
\end{center}
\begin{quote}

\baselineskip=16pt

Many models of supersymmetry breaking, in the context of either supergravity
or superstring theories,  predict the presence of 
particles with weak scale masses and Planck-suppressed
couplings. Typical examples are 
the scalar moduli  and the gravitino.
Excessive production of such particles in the early Universe
destroys the successful predictions of nucleosynthesis.
In particular, the thermal production of these relics 
after inflation leads to a bound on the 
reheating temperature, $T_{RH}\simlt 10^9$ GeV.  In this paper we 
show that the non-thermal generation of  these dangerous relics
may be much more efficient than the thermal production after inflation.
Scalar moduli fields
may be copiously  created  by the classical gravitational effects on the vacuum 
state. Consequently,
the new upper bound on the reheating temperature is shown to be, in some cases,
as low as  100 GeV. We also study the non-thermal production of gravitinos
in the early Universe, which can be  extremely efficient and  
overcome the thermal production by several orders of magnitude,
in  realistic supersymmetric inflationary models.

\end{quote}
\vskip1.cm
\begin{flushleft}
July  1999 \\
\end{flushleft}

\end{titlepage}
\setcounter{footnote}{0}
\setcounter{page}{0}
\newpage
%
\baselineskip=18pt
\noindent

\section{Introduction}

One of the problems facing Planck-scale physics is 
the lack of predictivity for low-energy phenomena.
Nonetheless, some information on high-energy 
physics  may be inferred indirectly through 
its effects on the cosmological evolution 
of the early Universe. If we require  that  the successful predictions of
big-bang
nucleosynthesis are not significantly modified
and that the energy density of stable particles 
does not overclose the Universe,  severe constraints are imposed 
on the properties of a 
large class of supersymmetric theories.

In  $N=1$ supergravity models 
\cite{sugra}, supersymmetry 
is broken in some hidden
 sector and gravitational-strength interactions communicate
the breaking down to 
the visible sector. In these models 
there often exist scalar and fermionic fields with masses of the 
order of the weak scale and 
gravitational-strength couplings to 
ordinary matter. In the following, we will generically  refer to  them 
as
{\it gravitational relics} $X$. 
If produced in the early Universe, such quanta will  behave 
like nonrelativistic matter and decay at very late times, eventually 
dominating the energy of the
 Universe until it is too late for nucleosynthesis to 
occur (in the case of scalar fields,
long-lived coherent oscillations with
large amplitudes of the zero mode can  pose the same problem).
Typical examples of gravitational relics 
are the spin-3/2 gravitino -- the supersymmetric partner of the graviton 
-- and the moduli fields -- the quanta of the scalar fields which
parametrize  supersymmetric flat directions in moduli space and seem almost 
ubiquitous in string theory. 

In string models massless 
fields exist in all known ground states and parametrize the continuous
vacuum degeneracies characteristic of supersymmetric theories. 
Possible examples of gravitational relics in string theory are the 
dilaton whose vacuum expectation value parametrizes the strength of the 
gauge forces, and the massless scalar (and fermionic superpartner)
gauge singlets parametrizing the size of the 
compactified dimensions.
In general,
these fields are massless to 
all orders in perturbation theory in the exact supersymmetric limit
and become massive either because of non-perturbative effects or because
of supersymmetry-breaking contributions. In the usual scenarios in which
supersymmetry breaking occurs at an intermediate scale $M_S = {\cal O}
(\sqrt{\mpl M_W})$, moduli acquire masses of the order of the weak scale
and, because of their long lifetime, 
pose a serious cosmological problem~\cite{moduli}. The problem
becomes even more severe in models in which supersymmetry is broken at
lower scales, since the moduli are lighter than $M_W$ and their lifetime
becomes exceedingly long. On the other hand, if supersymmetry breaking
occurs at a scale larger than $M_S$ and its effect on the observable sector
is screened, the cosmological problem of gravitational relics is relaxed.
In particular, this is the case of the recently-proposed models
with anomaly-mediated supersymmetry-breaking~\cite{anom}, in which moduli
and gravitinos can avoid cosmological difficulties~\cite{anomcos}.

Gravitational relics $X$ can be produced in the early Universe  because
of
thermal scatterings in the plasma.  The slow decay rate of the $X$-particles
is the essential source of cosmological problems because 
the  decay products of these relics  will destroy the $^4$He and 
D nuclei by photodissociation, 
and thus successful nucleosynthesis predictions \cite{nucleo,ellis}. The most 
stringent bound 
comes from the resulting 
overproduction of D $+$ $^3$He; this requires that the 
$X$-abundance 
 relative to the entropy density at the time of reheating after 
inflation should satisfy~\cite{kaw}
\be
\label{lll}
\frac{n_{X}}{s}\simlt 10^{-12}.
\ee
(The
exact bound depends upon the $X$ mass, and here we have assumed $M_X\sim$~TeV.)

Neglecting any initial number density, the Boltzmann
equation for the number density of gravitational relics  during 
the thermalization stage after inflation is
\be
\label{l}
\frac{d n_{X}}{dt}+3H n_{X}\simeq\langle \sigma_{{\rm tot}}
 |v|\rangle n^2_{{\rm light}},
\ee
where $\sigma_{{\rm tot}}\propto 1/\mpl^2$ is the total production 
cross section  and 
$n_{{\rm light}}\sim T^3$ 
represents the number density of light particles in the thermal bath. 
The  number  density $n_X$
at thermalization is obtained by solving \eq{l}, 
\be
\label{ll}
\frac{n_{X}}{s}\simeq 10^{-2}\:\frac{T_{RH}}{\mpl}.
\ee
Comparing eqs.~(\ref{lll}) and (\ref{ll}), one obtains an upper 
bound on the reheating temperature after inflation \cite{ellis} 
\be
\label{bound}
T_{RH}\simlt (10^{8}-10^{9})\: {\rm GeV}.
\label{trhlim}
\ee
If $T_{RH}\sim M_{{\rm GUT}}\sim 10^{16}$ GeV, gravitational relics 
are inconsistent with
nucleosynthesis. Moreover, if the initial state after inflation is free 
from gravitational relics, the reheating temperature in \eq{trhlim}
is too low to allow for the creation of  superheavy GUT bosons that can
eventually produce the baryon asymmetry \cite{review}.

The goal of this paper is to point out that gravitational relics can
be created   with 
dangerously  large   abundances  by  {\it non-thermal} effects. 

One possibility is represented by the non-thermal effect 
of the
classical gravitational background on the vacuum state during the 
evolution of the Universe. The particle creation 
mechanism is similar to
the inflationary generation of gravitational perturbations that seed
the formation of large scale structures.  However, the quantum
generation of energy density fluctuations from inflation is associated
with the inflaton field which dominated the mass density of the
Universe, and not a generic, sub-dominant scalar or fermionic field.
Gravitational particle creation has recently been employed to generate
non-thermal populations of
very massive particles~\cite{superheavy,kt98}. In 
particular, the 
desired abundance of superheavy dark matter particles may be
generated during the transition from the inflationary phase to a
matter/radiation dominated phase as the result of the expansion of the
background spacetime acting on vacuum quantum fluctuations of the dark
matter field.  Contrary to
the generation of density perturbations,  this  mechanism contributes 
to the homogeneous
background energy density that drives the cosmic expansion, and is
essentially the familiar ``particle production'' effect of
relativistic field theory in external fields.

Another possibility is represented 
by the non-thermal effects occuring right after 
inflation because of the rapid oscillations of the inflaton field(s).
Gravitino production is an interesting example of this kind.

The paper is organized as follows. In sect.~2 we 
show that the gravitational production of scalar moduli which 
are minimally coupled to gravity
may be much more efficient than  the one activated by thermal scatterings
during reheating. Our findings lead  to extremely stringent constraints on 
the reheating temperature  after inflation. 
In sect.~3 we study the non-thermal production of gravitinos
in the early Universe, showing  that the helicity-1/2 part of the 
gravitino can be efficiently excited during the evolution of the Universe. 
This leads to a copious and dangerous generation of gravitinos in 
realistic supersymmetric models of inflation.
Previous studies~\cite{porc} of non-thermal production of gravitinos
have considered only the subleading effects of the helicity-3/2 components.
However, a thorough study of the complete gravitino equations has
very recently appeared~\cite{g}. Although the derivation of the gravitino
equations presented in our paper is different from that of ref.~\cite{g},
our final results agree fully with the analysis of ref.~\cite{g}.
Finally, sect.~4 contains our conclusions.

\section{Gravitational Production of Scalar Moduli}

Let us describe here the basic physics underlying the mechanism of
gravitational scalar particle production.

We start by canonically quantizing the  action of a generic scalar 
massive field $X$ which, at the end, will be   identified with a modulus
field.  In the system of coordinates where the line element is 
given by   $ds^2= dt^2- a^2(t) d{\vec x}^2$, 
the action  is
\begin{equation}
S=\int dt \int d^3\!x\, \frac{a^3}{2}\left[ \left( \frac{dX}{dt}\right)^2
- \frac{({\vec \nabla }
X)^2}{a^2} - M_X^2 X^2 - \xi R X^2 \right] ,
\end{equation}
where $R$ is the Ricci scalar.  After transforming to conformal time
coordinate, where the line element is
$ds^2=a^2(\eta) (d\eta^2 - d{\vec x}^2)$,
we use the field mode expansion
\begin{equation}
X=\int \frac{d^3\!k}{(2 \pi)^{3/2} a(\eta)} \left[a_k h_k(\eta) e^{i
{\vec k} \cdot {\vec x}} + a_k^\dagger h_k^*(\eta) e^{-i 
{\vec k} \cdot {\vec x}}  \right] .
\end{equation}
Since the creation and annihilation operators obey the
commutator relations 
$[a_{k_1}, a_{k_2}^\dagger] = \delta^{(3)}({\vec k}_1 -{\vec
k}_2)$, we obtain the normalization condition $h_k {\dot h}_k^{*} - {\dot h}_k
h_k^* = i$  (henceforth, the dots over functions denote
derivatives with respect to
$\eta$).  The resulting mode equation is
\begin{equation}
\label{fun}
{\ddot h}_k(\eta) + \omega_k^2(\eta) h_k(\eta) = 0,
\end{equation}
where 
\begin{equation}
\label{klein}
\omega_k^2= k^2 + M_X^2 a^2 + (6 \xi - 1) \frac{\ddot a}{a} \ .
\end{equation}
The parameter $\xi$ is 1/6 for conformal coupling and 0 for minimal
coupling.  
The equation of massless conformally coupled quanta reduces
to the equation in flat space-time and, in this case, there is no particle
production. When $M_X\ne 0$ or $\xi \ne 1/6$,
conformal invariance is broken and particles are 
created. In the following we will be especially interested in the case
of scalar particles with non-conformal coupling. In particular, the case of
minimal coupling ($\xi =0$) is certainly of physical interest.
Just to give one example, in the toroidal compactification of the type 
IIB string theory, the modulus field describing the volume of the 
internal compactified dimensions is minimally coupled to gravity in the 
four dimensional effective action.

The differential equation (\ref{fun}) can be solved once the
boundary conditions are supplied.  Since the annihilation operator is
just a coefficient of an expansion in a particular basis, fixing the
boundary conditions is equivalent to fixing the vacuum.
We choose the  initial conditions as 
\begin{equation}
h_k(\eta_0)=\omega_k^{-1/2},\;\:\; {\dot h}_k(\eta_0)=-i\omega h_k(\eta_0),
\end{equation}
corresponding to a vanishing particle density at $\eta =\eta_0$.
To obtain the number density of the produced particles, we 
perform a Bogolyubov transformation from the vacuum mode solution with
the boundary condition at $\eta=\eta_0$  into the one with the boundary
condition at $\eta= \eta_1$ (any later time at which the particles are
no longer being created).  Defining the
Bogolyubov transformation as $ 
h_k^{\eta_1}(\eta)= \alpha_k h_k^{\eta_0}(\eta) + \beta_k h_k^{*
\eta_0}(\eta)$ (the superscripts denote where the boundary condition
is set), we obtain the following number density of produced particles:
\begin{equation}
n_X(\eta_1) =\frac{1}{2\pi^2 a^3(\eta_1)}\: \int_0^{\infty} \:dk\:k^2 
|\beta_k|^2 .
\end{equation} 
One should note that the number operator is defined at $\eta_1$
while the quantum state (approximated to be the vacuum state) defined
at $\eta_0$ does not change in time in the Heisenberg representation.

Let us now discuss the structure of the mass term of the $X$ scalar particle.
The flat directions in the supersymmetric potential, corresponding to
massless $X$ particles, are lifted by supersymmetry breaking and 
nonrenormalizable effects in the superpotential. An important observation 
for us is 
that in the early Universe global supersymmetry is broken (for instance 
because of the false vacuum energy density of the
inflaton). In supergravity theories, supersymmetry breaking is
transmitted by gravitational interactions and the supersymmetry-breaking  mass 
squared is naturally $C_H H^2$, where $H$ is the Hubble parameter and
$C_H={\cal O}(1)$. To illustrate 
this effect, consider a term in the
K\"{a}hler potential of the form 
\begin{equation}
\label{ch}
\delta K=-C_H\:\int\: d^4\:\theta \frac{1}{\mpl^2}I^\dagger IX^\dagger 
X,
\end{equation} 
where $I$ is the field which dominates the energy density $\rho$ of the 
Universe, that is $\rho\simeq \langle 
\int d^4\theta I^\dagger I\rangle$. During inflation, $I$ is 
identified with the inflaton field and $\rho=V(I)=3 H^2\mpl^2$. The 
term (\ref{ch}) therefore provides an effective $X$ mass 
$M^2_{X}=3C_H H^2$. The same mass term will appear during 
the coherent oscillations of the inflaton field $I$ after inflation. 
This example can be easily generalized, and the resulting 
potential of the scalar moduli fields during inflation is of the form  $
V=H^2\mpl^2\:{\cal V}(|X|/\mpl)$. We conclude that moduli quanta 
will be gravitationally produced in the early Universe with  a 
mass squared  of the form
\begin{equation}
 M_X^2\simeq m_X^2 +C_HH^2.
\end{equation} 
Here $m_X$ accounts for  the mass term generated by any  possible 
source of supersymmetry breaking whose $F$-term is not dominating the 
energy density
of the Universe during inflation, but persists in the zero-temperature
limit. For this reason, we expect $m_X$ to be of the order of TeV, the
present effective supersymmetry-breaking scale. 
It is important to bear in mind -- though -- that the case 
$C_H=0$ is certainly a possibility. Indeed, 
in supergravity models which possess a Heisenberg symmetry, supersymmetry 
breaking makes no contribution to scalar masses, leaving supersymmetric 
flat directions unmodified at tree-level.  No-scale supergravity of the 
$SU(N,1)$ type and the untwisted sectors from orbifold compactifications 
are special cases of this general set of models. 

We have numerically integrated \eq{fun}, in presence of an inflaton with
quadratic potential, up to the epoch when the $X$ field
starts oscillating on all scales and the field fluctuations
are transformed into particles which redshift as $a^{-3}$ thereafter.
This procedure determines a well-defined quantity, the 
total number of particles 
$n_X a^3$, which is constant at late times and 
is easy to compute numerically. However, the calculation is 
technically difficult in  the regime of  very small masses. In 
this regime a better strategy is represented by  adopting  
observables which are  independent of $m_X$
at small $m_X$.
In this limit -- and $C_H \rightarrow 0$ --  the present-day energy 
density of produced
$X$-particles is an appropriate quantity, since it
is independent of $m_X$ (see ref.~\cite{kt98} 
for a complete discussion  of this point). Therefore, 
for small masses, the number density $n_X$ is better inferred from 
the ratio $\rho_X/m_X$.  On the other hand, we expect that 
at $C_H \simgt 1$ the number density
of produced $X$-particles will be independent of $m_X$, since the field 
fluctuations will enter the  oscillating
regime at all scales at the epoch in which $M_X^2 \gg m_X^2$. 
Therefore,  in the limit of relatively
large $C_H$, it is more convenient to use $n_X$.

\begin{figure}
\leavevmode\epsfysize=8cm \epsfbox{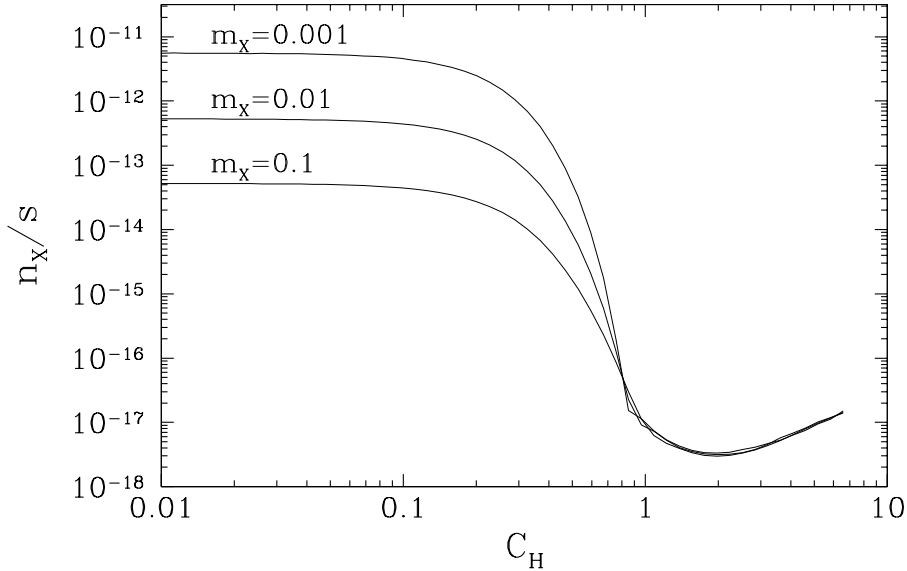}
\caption{The ratio $n_X/s$ as a function of $C_H$
for scalar moduli particles $X$
minimally coupled to gravity  and with squared masses $M_X^2=m_X^2+C_HH^2$.
In the figure
$m_X$ is expressed in units of the inflaton mass
$m_I$, and  $T_{RH}=10^9$ GeV, $m_I=10^{13}$ GeV.}
\label{fig:nX_mod}
\end{figure}

In our numerical results, 
we have  normalized $n_X$ by the entropy density. In 
doing so,  we have assumed that the reheating temperature after inflation is 
$T_{RH} = 10^9$ GeV and the mass of the inflaton field is $m_I = 10^{13}$ GeV. 
The results for scalar fields with minimal
coupling ($\xi=0$) are summarized in fig.~\ref{fig:nX_mod}. This 
figure  shows the limiting
behaviour of $n_X/s$ as discussed above, which allows to extract 
the relevant physical 
quantities at arbitrarily small $m_X$.
In particular, notice that for small 
$C_H$, the ratio $n_X/s$ scales like $m_X^{-1}$.
If the reader wishes to adopt different values for the reheating temperature
or the inflaton mass, the data in fig.~\ref{fig:nX_mod}
have to be multiplied by the ratio
\begin{equation} 
\left(\frac{T_{RH}}{10^9 {\rm ~GeV}}\right) \left(\frac{m_I}{10^{13} {\rm 
~GeV}}\right).
\end{equation}
Furthermore, if one wants to include the effect of
significant entropy release at some late time after 
reheating, the data in fig.~\ref{fig:nX_mod} have to be divided by the
amount of entropy increase in the comoving volume, 
$\gamma \equiv s_{\rm final}/s_{\rm initial}$. 

{}From fig.~\ref{fig:nX_mod} we infer that scalar moduli 
quanta are very efficiently produced by gravitational effects. 
In particular, for the realistic case $m_X\sim m_{3/2}
\sim 1$ TeV (or $m_X/m_I\simeq 10^{-10}$), the ratio 
$n_X/s$ turns out to be extremely large for small $C_H$
\begin{equation}
\frac{n_X}{s}\simeq 5\times 10^{-5}.
\end{equation} 
This result is seven  orders of magnitudes above the limit in \eq{lll}. 
In this case, the non-thermal production is so efficient 
that the thermal scatterings during reheating become 
completely irrelevant and the upper bound on the reheating 
temperature in order to get $n_X/s\simlt 10^{-12}$ becomes as low as 100 GeV.
Only when the parameter $C_H$ approaches unity, regardless of the value of
$m_X$, does the number density of $X$ particles 
become acceptably small.

Another way of presenting the bound in \eq{lll} is to compute  
the energy density in scalar moduli particles  normalized to the 
critical density $\rho_c$
in the Universe, $\Omega_X=\rho_X/\rho_c$. Were these particles stable,  
the parameter $\Omega_X$ would be
\be
\Omega_X h^2\simeq 4\times 10^{11} \left(\frac{m_X}{{\rm 
TeV}}\right)\left(\frac{n_X}{s}\right),
\ee
where $h$ is the Hubble rate in units of 100 km sec$^{-1}$ Mpc$^{-1}$.
Therefore the bound in \eq{lll} can be recast into the form
\be
\Omega_X h^2\simlt 4\times 10^{-1}\left(\frac{m_X}{{\rm TeV}}\right).
\ee 
The parameter $\Omega_X$ is shown in fig. 2.
Again, this figure was calculated assuming $T_{RH} = 10^9$ GeV 
and $m_I = 10^{13}$ GeV. 
If one wishes to adopt different values of $T_{RH}$ and $m_I$ or to take
into account some entropy release,
the data in this figure have to be
multiplied by the factor
\begin{equation} 
\gamma^{-1}\left(\frac{T_{RH}}{10^9 {\rm ~GeV}}\right) 
\left(\frac{m_I}{10^{13} {\rm ~GeV}}\right)^2.
\end{equation}

\begin{figure}
\leavevmode\epsfysize=8cm \epsfbox{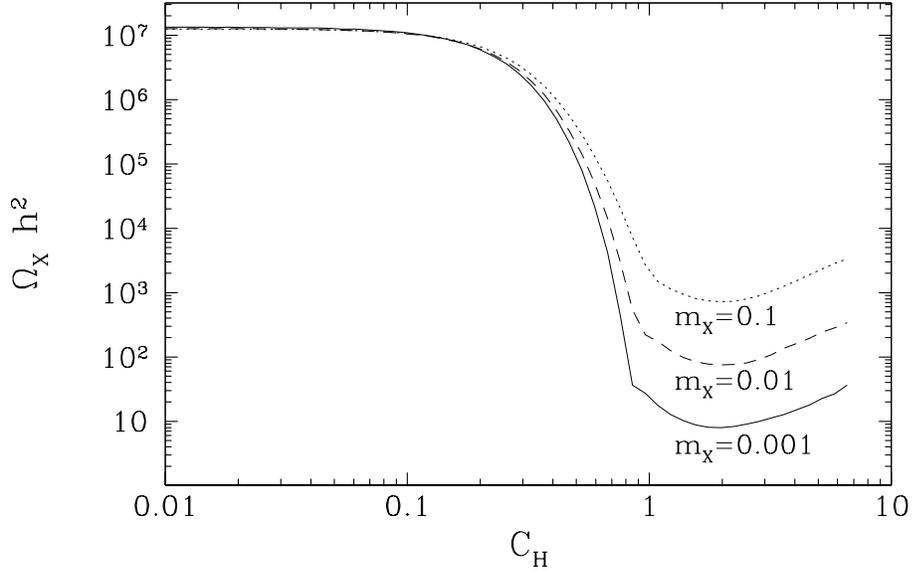}
\caption{Ratio of the energy density in scalar moduli particles 
minimally coupled to gravity to the critical density,
as a function of $C_H$. Conventions are the same as in fig. 1.}
\label{fig:rhoX_mod}
\end{figure}

\begin{figure}
\leavevmode\epsfysize=8cm \epsfbox{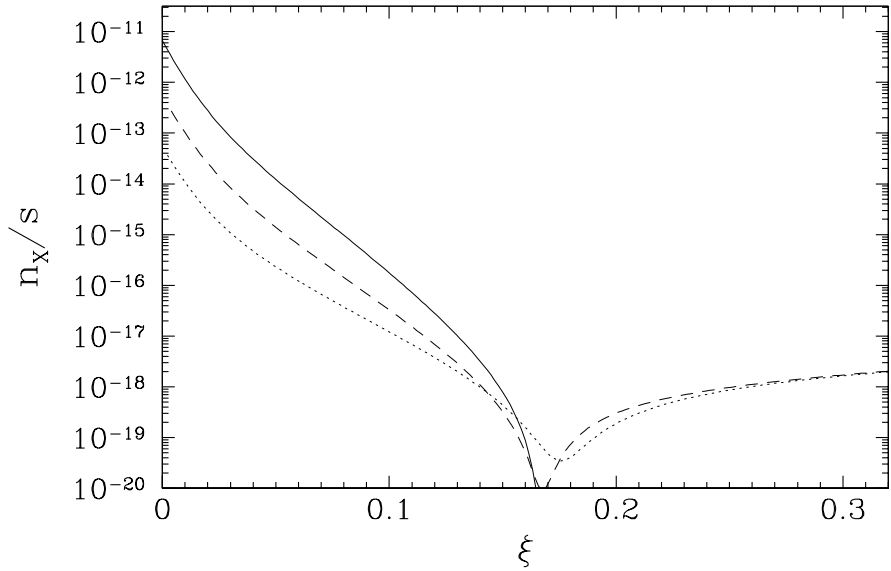}
\caption{The ratio $n_X/s$ of scalar moduli particles 
as a function of the parameter $\xi$, for 
$C_H=0$. The solid, long-dashed and short-dashed lines correspond to 
$m_X/m_I=0.001$, $m_X/m_I=0.01$ and $m_X/m_I=0.1$, respectively.}
\label{fig:xiscan_n}
\end{figure}

\begin{figure}
\leavevmode\epsfysize=8cm \epsfbox{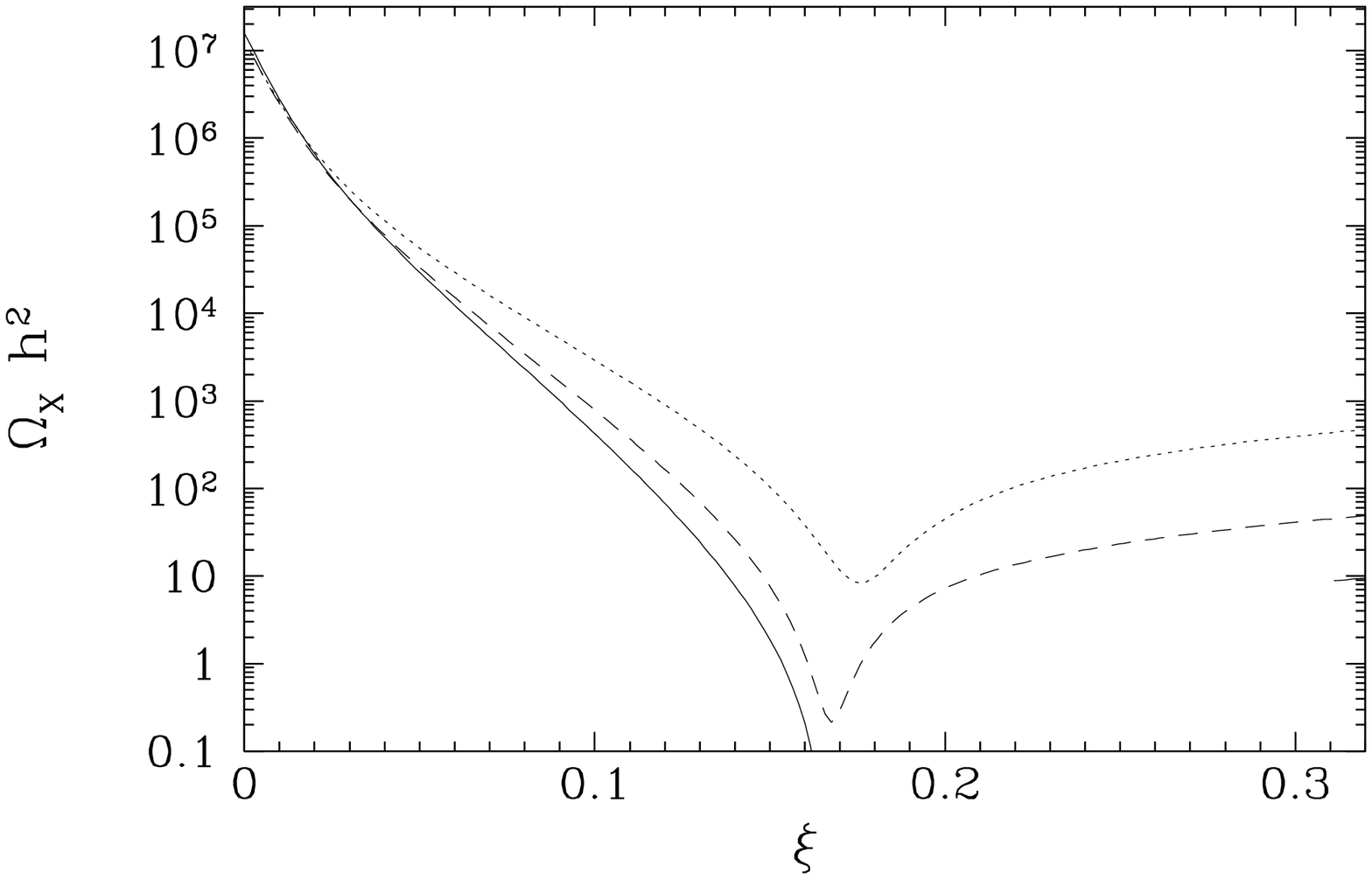}
\caption{Ratio of the energy density in scalar moduli particles 
to the critical density
as a function of $\xi$, for $C_H=0$. The solid, long-dashed and 
short-dashed lines correspond to 
$m_X/m_I=0.001$, $m_X/m_I=0.01$ and $m_X/m_I=0.1$, respectively.}
\label{fig:xiscan_omega}
\end{figure}

Figures 3 and 4 represent the ratio $n_X/s$ and $\Omega_Xh^2$
as a function of $\xi$ for $C_H=0$. From the data one can infer that 
scalar moduli particle production is extremely dangerous unless 
$\xi$ is very close to the conformal value $\xi=1/6$, where particle 
production shuts off for small
$m_X$. 

Our findings are also relevant for the idea  of enhanced symmetries 
and the ground state of string theory \cite{dine}. We know that there are 
often points in the moduli space with maximally enhanced symmetry, 
{\it i.e.} points where all of the moduli are charged under some symmetry 
(which may be continuous or discrete).  These moduli 
 do not  suffer from the cosmological moduli problem generated by 
their large coherent oscillations.
For ordinary moduli,  finite temperature and/or curvature effects 
(say during inflation) are
likely to leave the Universe in a state which does not correspond to the 
present  minimum of the moduli potential. The Universe remains in this 
state until the Hubble parameter is of order of the curvature of the potential, 
after which the system oscillates.  The moduli typically dominate the energy 
density of the Universe when they decay, leading to catastrophic consequences. 
However, if the vacuum is a point of maximally enhanced symmetry, it is quite 
natural for the Universe to start out in this state.
For example, during inflation, even though the potential for the moduli
is modified, it can be expected that the system
remains in the symmetric state.
Finite temperature effects also tend
to prefer states of higher symmetry \cite{dine}.  However, this selection rule
does not prevent the scalar quanta from being copiously produced 
by the gravitational effects, unless the effective mass of the 
scalar excitations
around the point of enhanced symmetry is much larger than the 
Hubble rate during
inflation. Therefore, gravitational 
production of these special moduli fields has to be checked case by case, in
order to assess the viability of a model.


\section{Non-Thermal Production of Gravitinos}

In this section we compute the non-thermal production of gravitinos in 
time-dependent gravitational backgrounds. Before solving the relevant
equations of motion in curved space, it is useful to recall the known results
of gravitino propagation in flat space.

The free propagation of massive gravitinos in Minkowski space is described
by four Majorana spinors $\psi^a$, satisfying
the Rarita-Schwinger equation
\beq
R^a\equiv 
\epsilon^{abcd} \gamma_5 \gamma_b \partial_c \psi_d +2m \sigma^{ab}\psi_b=0 .
\label{rarflat}
\eeq
Here flat space indices are denoted by Latin letters and are
contracted by the metric $\eta ={\rm diag}(+,-,-,-)$; we also use the
convention
\beq
\epsilon^{0123}=+1, ~~~~\sigma^{ab}=\frac{1}{4}\left[ \gamma_a , \gamma_b 
\right] .
\eeq
The mass term in \eq{rarflat} explicitly breaks supersymmetry. However, 
we are implicitly assuming that supersymmetry is spontaneously broken and 
\eq{rarflat} follows from choosing 
a supersymmetric gauge such that the Goldstino field
is fully eliminated from the Lagrangian. This choice is analogous to the unitary
gauge in Yang-Mills gauge theories. 

Using the identity
\beq
\epsilon^{abcd}\gamma_5 \gamma_b =\frac{i}{2} \left( \gamma^a \gamma^b
\gamma^c - \gamma^c \gamma^b \gamma^a \right) ,
\label{ident}
\eeq
\eq{rarflat} becomes
\beq
R^a=\frac{i}{2} \left( \gamma^a ~\dslash ~\gamma \cdot \psi -\gamma^b ~\dslash
~\gamma^a ~\psi_b \right) +m \left( \gamma^a ~\gamma \cdot \psi
-\psi^a \right)
=0 .
\label{rarsimp}
\eeq
{}From \eq{rarsimp} we obtain the following two constraints
\bea
\gamma \cdot R &=&2i \left( \dslash ~\gamma \cdot \psi -\partial \cdot \psi 
\right) +3m ~\gamma \cdot \psi =0, \label{con1} \\
\partial \cdot R &=& m \left( \dslash ~\gamma \cdot \psi -\partial \cdot \psi 
\right) =0. \label{con2}
\eea
For $m\ne 0$, these constraints imply the conditions $\gamma \cdot \psi =0$ and 
$\partial \cdot \psi =0$, which eliminate two spinors out of $\psi^a$,
leaving the appropriate degrees of freedom for the propagation of the
spin $\pm 3/2$ and $\pm 1/2$ components. Replacing 
eqs.~(\ref{con1})--(\ref{con2}) in \eq{rarsimp}, we obtain that the propagation 
of physical
states obeys the ordinary Dirac equation
\beq
\left( i \dslash -m \right) \psi^a =0.
\eeq

For $m=0$, \eq{rarflat} possesses a gauge symmetry $\delta_\epsilon \psi_a
=\partial_a \epsilon$, and we can choose a gauge fixing such that
$\gamma \cdot \psi =0$.
Then, \eq{con1} leads to $\partial \cdot \psi =0$. However, the choice
$\gamma \cdot \psi =0$ still allows
gauge transformation subject to $\dslash \epsilon =0$, 
and a further condition is required to fully fix the gauge. This condition
eliminates an additional fermionic component, leaving only the $\pm 3/2$ spin 
states as physical particles. Notice the complete analogy with the
case of spin-one particles, in which the Lorentz condition $\partial \cdot A
=0$
follows from the equation of motion in the massive case and it is a gauge
choice in the massless case. Also, for a massless spin-one particle, the
gauge fixing requires a further condition, which
eliminates the scalar degree of
freedom.

Let us turn now to the case of curved space, in which the gravitino equation
of motion becomes
\beq
R^\mu \equiv \epsilon^{\mu \nu \rho \sigma}\gamma_5 \hgam_\nu {\cal 
D}_\rho \psi_\sigma =0.
\label{rarcur}
\eeq
Greek letters denote space-time indices and Latin letters denote tangent-space
indices. Gamma matrices and the Levi-Civita symbol with curved indices are
defined by
\beq
\hgam^\mu \equiv e^\mu_a \gamma^a ,~~~~~~\epsilon^{\mu \nu \rho \sigma}
\equiv e e^\mu_a e^\nu_b e^\rho_c e^\sigma_d \epsilon^{abcd},
\eeq
where $e^\mu_a$ is the vierbein.
$\cal D_\mu$ is the covariant derivative with respect to the spinorial structure
modified to also reproduce the mass term
\beq
{\cal D}_\mu =D_\mu +\frac{i}{2} m \hgam_\mu ,~~~~D_\mu=\partial_\mu
+\frac{1}{2} \omega_{\mu ab} \sigma^{ab}.
\label{cov}
\eeq
The spin connection $\omega_{\mu ab}$ can be obtained (in first-order
formalism) by solving the supergravity
equation of motion where the gravitino, the vierbein,
and the spin connection are treated  as independent fields. One finds (see
{\it e.g.} ref.~\cite{casanova})
\bea
\omega_{\mu ab} &=&\frac{1}{2} \left( -C_{\mu ab} +C_{ab\mu}+C_{b\mu a}
\right), \\
{C^a}_{\mu \nu} &=&\partial_\mu e_\nu^a -\partial_\nu e_\mu^a -\frac{1}{2\mpl^2}
{\bar \psi}_\mu \gamma^a \psi_\nu, 
\label{torsion}
\eea
where ${\bar \psi}_\mu\equiv \psi^\dagger_\mu \gamma^0$.

For our cosmological considerations, we are interested in the case 
of spatially flat Friedmann-Robertson-Walker metrics, in which the line
element can be written as $ds^2=a^2(\eta)(d\eta^2-d{\vec x}^2)$. Here 
${\vec x}=(x^1,x^2,x^3)$ are comoving space coordinates,
$x^0\equiv \eta$ is the conformal time and $a$ is the scale factor, such that
$a^{-1}=d\eta /dt$.
Thus, the vierbein and the metric can be written
as $e^a_\mu =a\delta^a_\mu$, $e_a^\mu =a^{-1}\delta_a^\mu$, 
$g_{\mu \nu}=a^2\eta_{\mu \nu}$, $g^{\mu \nu}=a^{-2}\eta^{\mu \nu}$. In this
case, the covariant derivative in \eq{cov} reduces to
\beq
{\cal D}_\mu =\partial_\mu +\frac{\adot}{4a^2}\left(
\hgam_\mu \gamma^0 - \gamma^0 \hgam_\mu \right) +\frac{i}{2} m \hgam_\mu,
\label{covex}
\eeq
where the dot denotes the derivative with respect to conformal time.
Here we have dropped from \eq{torsion}
the torsion term, bilinear in the gravitino field. This term is crucial
for establishing the consistency of supergravity~\cite{deserzum}. 
However, for our considerations, it can be safely ignored 
since the number density of produced gravitinos is sufficiently small. 
During the cosmological epochs we will consider, both $m$ and $H$ are much
smaller than the Planck mass.

Using the identity in \eq{ident}, we can rewrite \eq{rarcur} in the 
Friedmann-Robertson-Walker metric as
\bea
&R^\mu&=i\left( \dslash \psi^\mu -\partial^\mu \hgam \cdot \psi +
\hgam^\mu \dslash \hgam \cdot \psi -\hgam^\mu \partial \cdot \psi \right), 
\nonumber \\
&+& i\frac{\adot}{2a} \left( 5 \hgam^0 \psi^\mu -3 \hgam^\mu \psi^0
+3\hgam^\mu \hgam^0 \hgam \cdot \psi -g^{\mu 0}\hgam \cdot \psi \right)
+m \left( \hgam^\mu \hgam \cdot \psi - \psi^\mu\right) =0,
\label{rarcas}
\eea
where $\dslash \equiv \hgam^\mu \partial_\mu$ and $\partial \cdot \psi
\equiv \partial^\mu \psi_\mu=\partial_\mu \psi^\mu +2\adot /a \psi^0$.
The condition ${\cal D}\cdot R =0$ gives the following algebraic 
constraint\footnote{The parameter $c$ can be 
expressed  in terms of the background 
energy density $\rho$ and  pressure $p$ using the Einstein equations 
\begin{equation}
c=\frac{\left(p-3m^2\mpl^2\right)\gamma^0+2i\dot{m}\mpl^2}{a\left(
\rho +3 m^2\mpl^2\right)}.
\nonumber
\end{equation} 
This expression agrees with the result found in ref.~\cite{g}.}:
\bea
\psi^0&=&c\sum_{i=1}^3\hgam_i\psi^i,
\label{condiz1}\\
c&=&\frac{1}{3a\left( \frac{\adot^2}{a^4}+m^2\right)}\left[ \left( -2
\frac{\addot}{a^3}+\frac{\adot^2}{a^4}-3m^2\right) \gamma^0 +2i\frac{\mdot}{a}
\right].
\eea
Here we have also considered the possibility that the gravitino mass is
time dependent, since in general it is determined by time-varying 
supersymmetry-breaking background fields. In the flat space-time limit 
with constant gravitino mass, one gets $c=-\gamma^0$ and recovers 
the gauge fixing condition $\gamma\cdot\psi=0$.

Since spatial translations generate an exact symmetry of space-time, it is 
convenient to expand the gravitino field in momentum modes, $\psi^\mu(\eta 
,\vec{x})\sim \int d^3 k e^{-i\vec{k}\cdot \vec{x}}\psi_{\vec{k}}^\mu (\eta)$.
In the following we will consider the equation of motion for a single momentum
mode and choose the coordinates such that $\vec{k}$ is along the $x^3$ 
direction. For simplicity, we drop the index $\vec{k}$ from 
$\psi_{\vec{k}}^\mu$. Because of the antisymmetric properties of the
Levi-Civita symbol, the equation $R^0=0$ 
does not contain time derivatives and therefore
describes an algebraic constraint on the gravitino momentum modes, which is 
given by
\bea
\psi^3&=&\left( d-\hgam^3\right) \left( \hgam_1 \psi^1+ \hgam_2 \psi^2\right),
\label{condiz2}
\\
d&=&\frac{k}{a^2\left( \frac{\adot^2}{a^4}+m^2\right)}\left( i\frac{\adot}{a^2}
\gamma^0 +m\right),
\eea
where $k\equiv |\vec{k}|$.
It is convenient to describe the remaining independent fields with two Majorana
spinors $\psi_{1/2}$ and $\psi_{3/2}$, defined such that
\bea
\psi^0 &=& \sqrt{\frac{2}{3}}~ c~\hgam_3~d~\hgam_1~ \psi_{1/2}, \label{psi0}\\
\psi^1 &=& \frac{1}{\sqrt{6}} ~\psi_{1/2}+ \frac{1}{\sqrt{2}}~ \psi_{3/2},\\
\psi^2 &=& \hgam^2\hgam_1\left( \frac{1}{\sqrt{6}} ~\psi_{1/2}-
 \frac{1}{\sqrt{2}}~ \psi_{3/2}\right),\\
\psi^3 &=& \sqrt{\frac{2}{3}}\left( d-\hgam^3\right) \hgam_1 ~\psi_{1/2}.
\label{psi3}
\eea
We can show that, in the flat limit and on mass-shell,  
$\psi_{1/2}$ and $\psi_{3/2}$ correspond to the $\pm 1/2$
and $\pm 3/2$ helicity states
by explicitly constructing the helicity 
projectors. Since the gravitino field is built from the direct product
of spin $1/2$ (spinorial indices) and spin $1$ (vectorial indices) states,
the helicity projectors can be decomposed using the appropriate 
Clebsch-Gordan coefficients
\bea
{\cal P}^\mu_{\pm 3/2} &=&P_{\pm 1/2} P^\mu_{\pm 1}, 
\label{pro12}\\
{\cal P}^\mu_{\pm 1/2} &=&\sqrt{\frac{1}{3}} P_{\mp 1/2}P^\mu_{\pm 1}
+\sqrt{\frac{2}{3}} P_{\pm 1/2}P^\mu_0.
\label{pro32}
\eea
The helicity projectors acting on spinorial ($P_{\pm 1/2}$)
and vectorial ($P^\mu_{\pm 1,0}$) indices
are 
\bea
&P_{\pm 1/2}=\frac{1}{2}\left( 1\pm i\gamma^1 \gamma^2\right),&\\
&P^\mu_{\pm 1}=\frac{1}{\sqrt{2}}(0,\mp 1,i,0), ~~~~P^\mu_0=\frac{1}{m}(k,0,0,
\sqrt{k^2+m^2})&
\eea
It is now 
easy to verify that, in the flat limit and on mass-shell, eqs.~(\ref{pro12})
and (\ref{pro32}) project $\psi^\mu$ onto $\psi_{1/2}$ and $\psi_{3/2}$,
respectively. One can also verify that the normalization chosen in
eqs.~(\ref{psi0})--(\ref{psi3}) insures that the Rarita-Schwinger Lagrangian
leads to canonical kinetic terms for the fields $\psi_{1/2}$ and 
$\psi_{3/2}$\footnote{Notice that the linear  combination 
\be
\sum_{i=1}^3
\hat{\gamma}_i\psi^i=\sqrt{\frac{2}{3}}\hgam_3 d \hgam_1 ~\psi_{1/2}
\nonumber
\ee
defines the $\pm 1/2$ helicity state 
adopted in ref. \cite{g}. The states $\psi_{1/2}$ 
and $\sum_i\hat{\gamma}_i\psi^i$ differ by
a time-dependent function.}.

The equations of motion for the fields $\psi_{1/2}$ and $\psi_{3/2}$ are
derived from \eq{rarcas}, 
\bea
&&\left[ i\gamma^0\partial_0 +i\frac{5\adot}{2a}\gamma^0 -ma
+k\gamma^3 \right] \psi_{3/2} =0,
\label{eqm32}\\
&&\left[ i\gamma^0\partial_0 +i\frac{5\adot}{2a}\gamma^0 -ma
+k\left( A+iB\gamma^0\right) \gamma^3 \right] \psi_{1/2} =0,
\label{eqm12}\\
&&A=\frac{1}{3\left( \frac{\adot^2}{a^4}+m^2\right)^2}\left[
2\frac{\addot}{a^3}\left(m^2-\frac{\adot^2}{a^4}\right) +
\frac{\adot^4}{a^8}-4m^2\frac{\adot^2}{a^4}+3m^4-4\frac{\adot}{a^3}\mdot 
m\right], \label{eqqa}\\
&& B=\frac{2m}{3\left( \frac{\adot^2}{a^4}+m^2\right)^2}\left[
2\frac{\addot \adot}{a^5}-\frac{\adot^3}{a^6}+3m^2\frac{\adot}{a^2}
+\frac{\mdot}{ma}\left(m^2-\frac{\adot^2}{a^4}\right) \right] . \label{eqqb}
\eea
With the field redefinition in eqs.~(\ref{psi0})--(\ref{psi3}), the
massive Rarita-Schwinger Lagrangian has been diagonalized.
The $\pm 3/2$-helicity states satisfy the same equation of motion as an
ordinary Dirac particle, except for a $5/2$ coefficient replacing the
usual $3/2$ in front of
the $\adot /a$ term. This coefficient is determined by the field scaling 
with $a$, and it  can be simply obtained by recalling that the
Lagrangian density should scale as an energy density, ${\cal L}\sim a^{-4}$.
Since $\partial_\mu \sim a^0$ and $\partial^\mu \sim a^{-2}$,
a simple inspection of the kinetic terms in the corresponding Lagrangians
shows that $\phi \sim a^{-1}$, $\psi \sim a^{-3/2}$, and $\psi^\mu \sim 
a^{-5/2}$
 for spin-0,
spin-1/2, and spin-3/2 particles, respectively.

The $\pm 1/2$-helicity states, on the other hand,  satisfy the more
complicated evolution described by \eq{eqm12}.
In the de Sitter limit ($\addot =2\adot^2/a$) with constant gravitino mass
($\mdot =0$), the coefficients in \eq{eqm12} become $A=\cos \theta$ and
$B=\sin \theta$, with $\tan (\theta /2) =\adot /(ma^2)$.
In general, however, the time dependendence of the gravitino mass $m$ is
related to the evolution of the scale factor. Let us consider the simple case
of a single chiral superfield $\Phi$ with minimal kinetic terms. The diagonal
time and space components of the Einstein equation become
\bea 
\frac{\adot^2}{a^4}&=& \frac{1}{3\mpl^2}\left[ V(\Phi)+
\left| \frac{d\Phi}{dt}\right|^2
\right] ,\label{ein1} \\
2\frac{\addot}{a^3}-\frac{\adot^2}{a^4}&=& \frac{1}{\mpl^2}\left[ V(\Phi)-
\left| \frac{d\Phi}{dt}\right|^2
\right] .\label{ein2} 
\eea
Using the expression for
the gravitino mass $m$ in terms
of the superpotential $W$,
\beq
m=e^{\frac{\Phi^\dagger \Phi}{2\mpl^2}}~\frac{|W(\Phi)|}{\mpl^2} ,
\eeq
we can write the scalar potential $V$ as
\beq
V=e^{\frac{\Phi^\dagger \Phi}{\mpl^2}}
\left[ \left| \partial_\Phi W +\frac{\Phi^\dagger W}{\mpl^2}
\right|^2 -\frac{3|W|^2}{\mpl^2} \right] = m^2 \mpl^2 \left[ \left|
\frac{\mdot \mpl}{am \frac{d\Phi}{dt}} \right|^2 -3 \right] . \label{scap}
\eeq
Replacing eqs.~(\ref{ein1}) and (\ref{ein2}) in \eq{scap}, we
obtain 
\beq
\mdot^2 = -\frac{\addot^2}{a^4}+\frac{\addot}{a}\left( \frac{\adot^2}{a^4}
-3m^2\right) +2\frac{\adot^4}{a^6} +6\frac{\adot^2}{a^2}m^2 . \label{magrel}
\eeq
When this expression for $\mdot$ is used in eqs.~(\ref{eqqa}) and 
(\ref{eqqb}), we obtain 
\beq
A^2+B^2=1.
\label{cazz}
\eeq
This condition plays an important r\^ole in the solution of the equation
for the $\pm 1/2$-helicity states.

To proceed, we explicitly write the momentum expansion for $\psi_{1/2}$ in
terms of creation and annihilation operators
\beq
\psi_{1/2}(\eta, \vec{x})= a^{-5/2} \int \frac{d^3k}{(2\pi)^{3/2}}
e^{-i\vec{k}\cdot \vec{x}}\sum_{r=1,2} \left[ u^r(\eta, \vec{k})a^r_k
+v^r(\eta, \vec{k})a^{r\dagger}_{-k} \right] ,
\eeq
where $v^r(\eta, \vec{k})= u^{rC}(\eta, -\vec{k})$. 
For simplicity, we now focus on a single polarization state and
omit the index $r$ in the following.
Choosing a gamma-matrix
representation such that 
\beq
\gamma^0 =\pmatrix{ \identity &0\cr 0&-\identity} ~~~~ 
\gamma^3=\pmatrix{ 0&\identity \cr -\identity &0},
\eeq
the spinor $u^r(\eta, \vec{k})$ satisfies the equation of motion (\ref{eqm12}),
{\it i.e.}
\beq
{\dot{u}}_\pm =\mp imau_\pm +k(iA\mp B) u_\mp ,
\label{eqmotu}
\eeq
where  $u^T=(u_+,u_-)$. The spinor $v(\eta, \vec{k})$ satisfies the same
equation of motion. The treatment of $\psi_{3/2}$ is analogous and it is 
obtained by just setting $A=1$ and $B=0$ in the previous equations.

We can now reduce the system (\ref{eqmotu}) of first-order 
differential equations into a second-order differential equation for 
the function $f=G^{-1/2}u_{+}$ with  $G\equiv A+iB$:
\begin{equation}
\ddot{f}+\left[
k^2 +m^2 a^2 +i(ma)^{\cdot}-
i\frac{\dot{G}}{G}ma
+\frac{\ddot{G}}{2G}-\frac{3}{4}\frac{{\dot{G}}^2}{G^2}
\right]f=0.
\label{nb}
\end{equation}
Here we have made use of the property $|G|=1$, see \eq{cazz}. Defining 
\begin{eqnarray}
G&\equiv& e^{2i\int^\eta d \eta ~\omega},\nonumber\\ 
\Omega&\equiv& \omega+ma,
\end{eqnarray}
\eq{nb} can be rewritten as
\begin{equation}
\ddot{f}+\left(k^2 +\Omega^2+i\dot{\Omega}\right)f=0.
\label{b}
\end{equation}
An analogous equation has been derived in ref.~\cite{g}. Equation~(\ref{b})
is identical to the
familiar equation for a spin-1/2 fermion in a 
time-varying background, if $\Omega$ is identified with $ma$.
To get a more transparent interpretation
of this frequency $\Omega$, it is useful to consider the limit 
$|\Phi|\ll \mpl$, as suggested in ref.~\cite{g}. 
Since $m\propto \mpl^{-2}$, in this limit the
function $G$ tends to the expression 
\begin{equation}
G=\frac{p-2i\dot{m}\mpl^2}{\rho},
\end{equation}
where $\rho=|d\Phi/dt|^2+|\partial_\Phi W|^2$ and 
$p=|d\Phi/dt|^2-|\partial_\Phi W|^2$. One can now verify that 
$G$ satisfies the following differential equation
\begin{equation}
 \frac{\dot{G}}{G}=-2i\frac{\partial^2 W}{\partial\Phi^2}.
\end{equation}
 Therefore, in the limit of $\mpl\rightarrow\infty$ and fixed $|\Phi|$,
$\omega\rightarrow -\partial_\Phi^2 W$ and  the frequency $\Omega$ 
of the oscillations corresponds to the superpotential mass parameter
of the Goldstino which 
is `eaten' by the gravitino when supersymmetry is broken. Therefore, 
\eq{b} describing the production of helicity-1/2 gravitinos 
in supergravity reduces, in the limit of 
$|\Phi|\ll \mpl$, to the equation describing the time evolution of 
the helicity-1/2 Goldstino in global supersymmetry and no
suppression by powers of $\mpl$ is present.  
The frequency of the oscillations $\Omega$ depends 
upon all the mass scales appearing in the 
problem, namely the Goldstino mass parameter
$\partial^2W/\partial\Phi^2$, the Hubble 
rate $H$ and the gravitino mass $m$. The production of the 
helicity-1/2 gravitino is expected to be dominated by the 
fastest time-varying mass scale in the problem.  

To compute the abundance of   helicity-1/2 gravitinos generated  during 
and after an inflationary stage in the early Universe, one needs to   
discriminate among    various  supersymmetric inflationary models 
\cite{revinf}. A crucial point to keep in mind 
is that  a generic supersymmetric inflationary stage dominated by an 
$F$-term has the problem that the flatness of the potential is 
spoiled by supergravity corrections or, in other words,
the slow-roll parameter 
$\eta=\mpl^2 V^{\prime\prime}/V$ gets contributions of
 order unity. In simple one chiral field models based on  
superpotentials of the type $W=m\Phi^2/2$ or higher powers in 
$\Phi$, $W\sim \Phi^n$,   supergravity corrections make inflation 
impossible to start. To construct a model of inflation in the context
 of  supergravity, one must either invoke 
 accidental 
cancellations \cite{linderiotto}, or a period of inflation dominated by a
$D$-term \cite{dterm},  or some  particular properties based  on string theory
\cite{gel}. 

Since the theory of production of helicity-1/2 gravitinos looks similar to the
case of helicity-1/2 fermions with a frequency $\Omega$, one
can use as a guide  the recent results obtained in the theory of generation of
Dirac fermions during preheating \cite{ferm}. During inflation, since the mass
scales present in the frequency $\Omega$ are approximately constant in time,
one does not expect a significant production of gravitinos (the number density
can be at most $n_{3/2}\sim H_I^3$, where $H_I$ is the value of the Hubble rate
during inflation). However, in the evolution of the Universe subsequent to
inflation, a large amount of gravitinos can be produced. During the inflaton
oscillations, the Fermi distribution function is rapidly saturated up to some
maximum value of the momentum $k$, {\it i.e} $n_k\simeq  1$ for $k\simlt
k_{{\rm max}}$ and it is zero otherwise.  The resulting number density is
therefore $n_{3/2}\sim k_{{\rm max}}^3$. The value of $k_{{\rm max}}$ is
expected to be roughly of the order of the inverse of the time-scale needed for
the change of the mass scales of the problem at hand. Let us give a realistic 
example.   Consider the superpotential\footnote{Our results on gravitino
production are based on a simple one chiral superfield model, but we expect
similar results to be valid in 
the case of a theory with  more than one superfield.}
\be
\label{sup}
W=  S(\kappa\bar{\phi} \phi- \mu^2),
\ee
where $\kappa$ is a dimensionless coupling of order unity \cite{sh}. Here,
$\phi$ and $\bar{\phi}$ are oppositely charged under all symmetries so that
their product is invariant. The canonically-normalized inflaton field is
$\Phi\equiv\sqrt 2|S|$. The superpotential (\ref{sup}) leads to hybrid
inflation. Indeed, for  $\Phi\gg \Phi_c=\mu /\sqrt{\kappa}$,
$\phi=\bar{\phi}=0$ and the potential reduces to $V=\mu^4$ plus supergravity
and logarithmic corrections \cite{linderiotto}. Therefore, in this regime the
Universe is trapped in the false vacuum and we have slow-roll inflation. The
scale $\mu$ is fixed to be around $10^{15}$ GeV to reproduce the observed
temperature anisotropy. Notice that in this period, the Goldstino mass
$\partial^2 W/\partial S^2$ is vanishing. 

When $\Phi=\Phi_c$, inflation ends because the false vacuum becomes unstable  
and the fields $\phi$ and $\bar{\phi}$ rapidly oscillate around  the minimum of
the potential at $\langle\bar{\phi}\phi\rangle=\mu^2/\kappa$, while the field
$\Phi$ rapidly oscillates around zero. The time-scale of the oscillations is
${\cal O}(\mu^{-1})$. The mass scales at the end of inflation change by an
amount of order of $\mu$ within a time-scale $\sim \mu^{-1}$. Therefore, one
expects $k_{{\rm max}}\sim \mu$   and $n_{3/2}\sim\alpha \mu^3$, where $\alpha$
summarizes the uncertainty in the estimate. If all the energy residing in the
vacuum during inflation is istantaneously transferred to radiation, the
reheating temperature would result to be $T_{RH}\sim \mu\sim 10^{15}$ GeV and
the ratio $n_{3/2}$ to the entropy density would be $n_{3/2}/s\sim \alpha$.
This is not a realistic situation -- though -- because such a high reheating
temperature is already ruled out by considerations about the gravitino
generation through thermal collisions, see the bound in \eq{bound}. In a more
realistic scenario in which reheating and thermalization occur sufficiently
late, the number density of gravitinos decreases as $a^{-3}$ -- $a$ being the
scale factor -- in the post-inflationary scenario, presumably characterized by
a matter-dominated Universe. If this is the case,   at reheating the final 
ratio $n_{3/2}$
to the entropy density is
\be
\frac{n_{3/2}}{s}\sim \alpha\frac{T_{RH}}{\mu}.
\ee
If $\alpha={\cal O}(1)$, this violates the bound in \eq{lll} by at least five
orders of magnitude even if $T_{RH}\sim 10^9$ GeV and imposes a stringent
upper bound on the reheating temperature
$T_{RH}\simlt 1$ TeV. Notice that  the non-thermal production is
about five orders of magnitude more efficient than the generation through
thermal scatterings during the reheating stage, irrespectively of the value of
$T_{RH}$. A similar result may be obtained in the case of $D$-term inflation. 

The ultimate reason for such a copious generation of gravitinos  is that the
system relaxes to the minimum in a time-scale   much shorter than the Hubble
time $\sim H_I^{-1}$, since the  frequency is 
set  by the height of the potential
$V^{1/4}\gg H_I$ during inflation. This is a common feature of realistic
models 
of supersymmetric inflation.

\section{Conclusions}

In conclusion, we have investigated the non-thermal production in the  
early Universe of particles  with mass in the TeV range and couplings  
to matter    suppressed by powers of  $\mpl$. If produced with too  
large abundances, the late decays of these gravitational relics may  
jeopardize the successful predictions of standard big-bang  
nucleosynthesis. We have shown that scalar moduli -- which are  
generically present in the mass spectrum of supergravity and string  
theories -- may populate the Universe in large amounts as a result of  
the expansion of the background space-time acting on the vacuum quantum  
fluctuations of the moduli fields. The resulting number to entropy  
density ratio depends  upon 
the way these scalar moduli are coupled to gravity and what is their  
effective mass during and after inflation. We have shown that the  
generation of these moduli fields  poses no problem only if they couple  
to gravity with $\xi$ close
to $1/6$, {\it i.e.} they are  conformally coupled, and if the  
effective mass squared $C_H H^2$ is comparable or larger
than the Hubble rate, {\it i.e}  
$C_H\simgt 1$. On the contrary, if $\xi\neq 1/6$ and $C_H\simlt 1$, the  
generation can be so
efficient that the standard predictions of 
nucleosynthesis are safe only if the  
final reheating temperature is as low as 100 GeV. This upper bound is  
considerably
smaller than the bound of about $10^9$ GeV obtained from 
considerations about the production of scalar moduli  
through thermal collisions during the reheat stage. 

In this paper,   
we have also studied the non-thermal generation of gravitinos in the  
early Universe. The
spin-3/2 part of the gravitino is excited only in very small amounts,   
as the resulting abundance is proportional to the small gravitino mass.
On the other hand,
the spin-1/2 part obeys the equation of motion of a normal helicity-1/2  
Dirac particle
in a background whose frequency is a combination of the different mass  
scales
at hand: the superpotential mass parameter
of the fermionic superpartner of the scalar field  
whose $F$-term breaks supersymmetry, the Hubble rate and the gravitino  
mass. Again, 
in most realistic models of supersymmetric inflation, the non-thermal   
production of gravitinos turns out to be much more efficient than their
thermal generation during the reheat stage after  
inflation.
We have found, for example, that the ratio $n_{3/2}/s$ for helicity-1/2
gravitinos in generic models of inflation  
is roughly given 
by $T_{RH}/V^{1/4}$, where $V^{1/4}\sim 10^{15}$ GeV is the height of  
the potential during inflation. This generically
leads to an overproduction of  
gravitinos. However, we should stress that the non-thermal gravitino
production is quite sensitive to the specific inflationary model. Therefore,
it will become an important ingredient in the search for realistic
supersymmetric models of inflation.

\vskip1cm
\centerline{\large\bf Acknowledgements}
\vskip 0.2cm

We would like to thank T. Banks, R. Brustein,  M. Dine, S.~Ferrara, L. Kofman,
A. Linde, D.H.~Lyth, M.~Porrati, and A.~Zaffaroni for many useful discussions.


\vskip1cm
\begin{thebibliography}{99}
%
\bibitem{sugra} For a review, see H.P. Nilles, Phys. Rep.
{\bf 110}, 1  (1984).

\bibitem{moduli} B. de Carlos, J.A. Casas, 
F. Quevedo, and E. Roulet, Phys. Lett. 
{\bf B318}, 447 (1993);\\
T. Banks, D. Kaplan, and A. Nelson, Phys. Rev. 
{\bf D49}, 779 (1994).

\bibitem{anom} L. Randall and R. Sundrum, hep-th/9810155;\\
G. Giudice, M.A. Luty, H. Murayama, and R. Rattazzi, JHEP {\bf 9812}, 
027 (1998).

\bibitem{anomcos} T. Gherghetta, G.F. Giudice, and J.D. Wells, hep-ph/9904378;\\
T. Moroi and L. Randall, hep-ph/9906527.
                 
\bibitem{nucleo} D. Lindley, Ap. J. {\bf 294}, 1  (1985);\\ J. Ellis {\it
et al.}, Nucl. Phys. {\bf 259}, 175 (1985);\\ S. Dimopoulos {\it et al.}, Nucl.
Phys. {\bf B311}, 699 (1988);\\ J. Ellis {\it et al.}, Nucl. Phys. 
{\bf B373}, 399 (1992). 

\bibitem{ellis} J. Ellis, A. Linde, and D. Nanopoulos, Phys. Lett. {\bf
B118}, 59 (1982); \\ D. Nanopoulos, K. Olive, and M. Srednicki, 
Phys. Lett. {\bf
B127}, 30 (1983);\\ J. Ellis, J. Kim, and D. Nanopoulos, 
Phys. Lett. {\bf B145}, 181 
(1984).


\bibitem{kaw} M. Kawasaki and T. Moroi, 
Prog. Theor. Phys. {\bf 93}, 879 (1995).
 
\bibitem{review} For a recent review, see A. Riotto and A. Trodden, {\it Recent
progress in baryogenesis}, hep-ph/9901362, to appear in Annual Review of Nuclear
and Particle Science. 




\bibitem{superheavy} D.J.H. Chung, E.W. Kolb, and  A. Riotto,
Phys. Rev. {\bf D59}, 023501 (1999);\\
D.J.H. Chung, E.W. Kolb and  A. Riotto,  Phys. Rev. Lett. {\bf 81}, 
4048 (1998);\\
D.J.H. Chung, E.W. Kolb and  A. Riotto, hep-ph/9809453; \\ 
D.J.H. Chung, E.W. Kolb and  A. Riotto, hep-ph/9810361.


\bibitem{kt98}  V. Kuzmin and I. Tkachev, JETP Lett. {\bf 68}, 271 (1998);\\
V. Kuzmin and I. Tkachev, Phys. Rev. {\bf D59}, 123006 (1999).


\bibitem{porc}  D.H. Lyth, D. Roberts, and M. Smith,  
Phys. Rev. {\bf D57}, 7120 (1998);  \\ 
A.L. Maroto and A. Mazumdar, hep-ph/9904206.
 



\bibitem{g} R. Kallosh, L. Kofman, A. Linde, and A. Van Proeyen, hep-th/9907124.

\bibitem{dine}   M. Dine,  Y. Nir, and   Y. Shadmi, 
Phys. Lett. {\bf B438}, 61 (1998).

\bibitem{casanova} P. van Nieuwenhuizen, Phys. Rep. {\bf 68}, 189 (1981).

\bibitem{deserzum} S. Deser and B. Zumino, Phys. Lett. {\bf 62B}, 335 (1976).


\bibitem{revinf}  For a review, see D.H. Lyth and  A. Riotto, Phys. Rep. {\bf
314}, 1 (1999).

\bibitem{linderiotto}  
A. D. Linde and A. Riotto,  
Phys. Rev. {\bf D56}, 1841 (1997). 


\bibitem{dterm} P. Binetruy and G. Dvali, Phys. Lett. {\bf B388}, 241 (1996);\\
E. Halyo, Phys. Lett. {\bf B387}, 43 (1996); \\ D.H. Lyth and A. Riotto,  Phys.
Lett. {\bf 412}, 28 (1997); \\ G. Dvali and A. Riotto,  Phys.
Lett. {\bf 417}, 20 (1998); \\ J.R. Espinosa, A. Riotto, and G.G.
Ross, Nucl. Phys. {\bf B531}, 461 (1998);\\ S.F. King 
and  A. Riotto, Phys. Lett. {\bf B442}, 68 (1998). 


\bibitem{gel} J.A. Casas, G.B. Gelmini, and A. Riotto, hep-ph/9903492.

\bibitem{ferm}J. Baacke, K. Heitmann, and C. Patzold, Phys. Rev. {\bf D58},
125013 (1998); \\ P. B. Greene and L. Kofman, Phys. Lett. {\bf B448}, 6 
(1999);\\
G.F. Giudice,  M. Peloso, A. Riotto, and I. Tkachev, hep-ph/9905242. 

\bibitem{sh} G. Dvali,  Q. Shafi, and R. Schaefer,  
Phys. Rev. Lett. {\bf 73}, 1886 (1994). 






\end{thebibliography}
\end{document}